\documentclass{llncs}
\usepackage{makeidx}  % allows for indexgeneration
\usepackage{graphicx}
\newcommand{\slug}{\hbox{\kern1.5pt\vrule width2.5pt height6pt depth1pt\kern1.5pt}}
\def\xskip{\hskip 7pt plus 3pt minus 4pt}
\def\sskip{\vskip 3pt plus 1pt minus 1pt}
\newdimen\algindent
\newif\ifitempar \itempartrue % normally true unless briefly set false
\def\algindentset#1{\setbox0\hbox{{\bf #1.\kern.25em}}\algindent=\wd0\relax}
\def\algbegin #1 #2{\algindentset{#21}\alg #1 #2} % when steps all have 1 digit
\def\aalgbegin #1 #2{\algindentset{#211}\alg #1 #2} % when 10 or more steps
\def\alg#1(#2). {\medbreak % Usage: \algbegin Algorithm A (algname). This...
  \noindent{\bf#1}({\it#2\/}).\xskip\ignorespaces}
\def\algstep#1.{\ifitempar\sskip\noindent\else\itempartrue
  \hskip-\parindent\fi
  \hbox to\algindent{\bf\hfil #1.\kern.25em}%
  \hangindent=\algindent\hangafter=1\ignorespaces}
\begin{document}
\title{Ranking Entities in the Age of Two Webs, \\An Application to Semantic Snippets}
\author{Mazen Alsarem\inst{1}, Pierre-Edouard Portier\inst{1}, Sylvie Calabretto\inst{1}, and \\Harald
Kosch\inst{2}}
\institute{Universit\'{e} de Lyon, CNRS\\
INSA de Lyon, LIRIS, UMR5205, F-69621, France
\and
Universit\"{a}t Passau\\
Innstr. 43, 94032 Passau, Germany}
%
%\author{Xxxxx Xxxxxxx\inst{1}, Xxxxxxxxxxxxxx Xxxxxxx\inst{1}, Xxxxxx Xxxxxxxxxx\inst{1}, and Xxxxxx
%Xxxxx\inst{2}}
%%
%\institute{Xxxxxxxxxx xx xxxxx xxxx\\
%Xxxx xx Xxxxx xxxxx xxxxxxxx xxxxxxxx xxxxxx
%\and
%Xxxxxxxxxxx Xxxxxx\\
%Xxxxxxx xxx xxxxx xxxxxxx Xxxxxxx}
\maketitle
\begin{abstract}
The advances of the Linked Open Data (LOD) initiative are giving rise to a more structured Web of data. Indeed, a few datasets act as hubs (e.g., DBpedia) connecting many other datasets. They also made possible new Web services for entity detection inside plain text (e.g., DBpedia Spotlight), thus allowing for new applications that can benefit from a combination of the Web of documents and the Web of data. To ease the emergence of these new applications, we propose a query-biased algorithm (LDRANK) for the ranking of web of data resources with associated textual data. Our algorithm combines link analysis with dimensionality reduction. We use crowdsourcing for building a publicly available and reusable dataset for the evaluation of query-biased ranking of Web of data resources detected in Web pages. We show that, on this dataset, LDRANK outperforms the state of the art. Finally, we use this algorithm for the construction of semantic snippets of which we evaluate the usefulness with a crowdsourcing-based approach.
\end{abstract}
\section{Introduction}
In this work, we introduce LDRANK (see section~\ref{ldrank}), an efficient query-biased and context-aware ranking algorithm that applies to the resources of a LOD graph. When combined with the automatic annotation of resources in Web pages (e.g. through DBpedia Spotlight~\cite{mendes2011dbpedia}), LDRANK offers the opportunity to build useful semantic snippet that can apply to any Web page regardless of its provenance (see section~\ref{ensen}). In this introduction, we provide the background information from which the necessity for this new algorithm will appear.

% Ranking on the Web of Documents VS Ranking on the Web of Data
%
On the web of documents links are indications of a relationship between information carried by the documents. Although these indications are coarse-grained, they revealed themselves as essential for the most-effective ranking algorithms (PageRank~\cite{page1999pagerank}, HITS~\cite{kleinberg1999authoritative}, SALSA~\cite{lempel2001salsa}).

On the web of data, links are fine-grained explicit relationships between resources (i.e., URI for things of the phenomenal world, be they mental or physical). The vast majority of the existing ranking strategies for the web of data (see~\cite{roasurvey} and~\cite{jindal2014review} for recent surveys) are relying on adaptations of PageRank. The modifications made to adapt the PageRank algorithm to the web of data are necessary due to the high heterogeneity of both the provenance of the datasets and the types of the relationships. Otherwise, there are also a few experiments with learning-to-rank approaches applied to the web of data (e.g.,~\cite{dali2012query}). These techniques depend on the availability of relevance judgments for training (although indirect measures of correlated quantities can sometimes be used, e.g. the number of visits agents made to a resource).

%DBpedia and the Linked Open Data (LOD) Initiative
%
In order to manage the aforementioned intrinsic heterogeneity of the web of data, the Linked Open Data (LOD) initiative promotes simple principles for publishing resources in a way conducive to a web of linked data with shared knowledge expressed in a common formalism (RDF) and accessible through a common interface (HTTP). As a key use-case, DBpedia has been used in conjunction with NLP strategies in order to associate resources with their surface forms in a text document. The main current applications for this use-case are: DBpedia Spotlight~\cite{mendes2011dbpedia}, AlchemyAPI\footnote{{\scriptsize\url{www.alchemyapi.com}~; \url{www.opencalais.com}~; \url{www.ontos.com}}} (similar to DBpedia Spotlight, but finds resources in various LOD datasets and thus includes a coreference resolution step), OpenCalais\footnotemark[\value{footnote}], SemanticAPI from Ontos\footnotemark[\value{footnote}], ZenCrowd~\cite{demartini2012zencrowd}\dots

% Ranking on the edge of two webs
%
In this context, we address the problem of ranking resources that come from the automatic annotation of a Web page selected by a web search engine in response to a user query. The main challenge is to make good use of the knowledge given by the query and the Web page's text in order to palliate the sparsity and heterogeneity of the graph of resources.
We propose an algorithm, LDRANK, and we compare it to other modified PageRank algorithms. Moreover, we apply it to the construction of semantic snippets\footnote{{\scriptsize\url{http://liris.cnrs.fr/drim/projects/ensen/}: live demo, source code, technical report, datasets}}. A snippet is an excerpt from a Web page determined at query-time and used to express how a Web page may be relevant to the query. A semantic snippet is meant to improve the process of matching the ranked Web pages presented within a Search Engine Result Page (SERP) with the user's mental model of her information need. It achieves this objective by making apparent the relationships existing between the information need and the more relevant resources present in the Web page.

In section~\ref{related-works} we introduce the related works about enhanced snippets for the web of documents and for the web of data. In section~\ref{dataset}, we describe the construction of a dataset for the evaluation query-biased entity ranking algorithms. In section~\ref{ldrank} we present the LDRANK algorithm and its evaluation. In section~\ref{ensen}, we introduce ENsEN, the software system we developed to provide semantic snippets. In section~\ref{user} we present the results of an evaluation of the usefulness of ENsEN.
\section{Related Works}\label{related-works}
We first mention works that generate snippets for native RDF documents. Ge~{\it et al.}~\cite{ge2012incorporating}, and Penin~{\it et al.}~\cite{penin2008snippet} focus on the generation of snippets for ontology search.  Bai~{\it et al.}~\cite{bai2008rdf} generate snippets for a semantic web search engine.

In~\cite{penin2008snippet}, the authors first identify a topic thanks to an off-line hierarchical clustering algorithm. Next, they compute a list of RDF sentences (i.e. sets of connected RDF statements) semantically close to the topic. Finally, they rank the selected RDF statements by considering both structural properties of the RDF graph and lexical features of the terms present in the ontology (by way of a Wordnet-based similarity measure).

In~\cite{ge2012incorporating}, the authors first transform the RDF graph into a term association graph in which each edge is associated with a set of RDF sentences. Their objective is to produce a compact representation of the relationships existing between the terms of the query. These relationships are to be found in the RDF graph. To do this, they decompose the term association graph into maximum r-radius components in order to avoid long distance relations between query terms. Next, they search sub-snippets in these components (i.e. connected subgraphs that link some of the query-terms).  Finally, they select some of the sub-snippets to form the final snippet.

In~\cite{bai2008rdf}, the authors first assign a topic to the RDF document (they use a property such as {\it p:primaryTopic} if it exists, otherwise they rely on a heuristic based on the comparison of the URI of the candidates topic-nodes with the text of the URL of the RDF document). Next they design a ranking algorithm for RDF statements. Particularly, they introduce the notions of {\it correlative} (e.g. \texttt{foaf:surname} and \texttt{foaf:family\_name}) and {\it exclusive} (e.g. \texttt{foaf:name} and \texttt{foaf:surname}) properties. Finally, they use this ranking algorithm to give the user a set of relationships between the query-related statements and the topic-related statements.

To sum up, we agree with Ge~{\it et al.}~\cite{ge2012incorporating} that the main benefit of possessing highly structured data from an RDF graph is the possibility to find non-trivial relationships among the query terms themselves, and also between the query terms and the main concepts of the document. Moreover, we agree with Penin~{\it et al.}~\cite{penin2008snippet} and Bai~{\it et al.}~\cite{bai2008rdf} about the necessity to design a ranking algorithm for RDF statements that considers both the structure of the RDF graph and lexical properties of the textual data. However, we find ourselves in an inverted situation with genuine text extracted from classical Web pages, and RDF graphs automatically generated from these Web pages.

Indeed, LOD resources can either come from: (i) a LOD dataset (e.g. by way of SPARQL queries), (ii) semantic annotations embedded in a Web page (i.e., by using RDFa, Microdata, or Microformats\footnote{{\scriptsize\url{www.w3.org/TR/xhtml-rdfa-primer/}~;
\url{microformats.org/}~; \url{www.w3.org/TR/microdata/}}}), or (iii) automatic association of resources with surface forms of the Web page by way of NLP strategies (e.g. DBpedia Spotlight~\cite{mendes2011dbpedia}, ZenCrowd~\cite{demartini2012zencrowd},\dots). Among the approaches that offer to enhance the snippets of a SERP by using the web of data~\cite{haas2011enhanced}~\cite{steiner2010google}, none rely on automatic annotation: they use only embedded annotations. Haas~{\it et al.}~\cite{haas2011enhanced} employed structured metadata (i.e. RDFa and several microformats) and information extraction techniques (i.e. handwritten or machine-learned wrappers designed for the top host names e.g., \url{en.wikipedia.org}, \url{youtube.com},\dots) to enhance the SERP with multimedia elements, key-value pairs and interactive features. By combining metadata authored by the documents' publishers with structured data extracted by ad-hoc wrappers designed for a few top host names, they are able to build enhanced snippets for many results of a SERP. They chose not to use the LOD graph to avoid the problem of the transfer of trust between the Web of documents and the Web of Data. Indeed, they argue that the quality of the editorial processes that produce the Web of Data from the Web of documents (e.g. the transformation from Wikipedia to DBPedia) cannot be controlled. Therefore, from their point of view, making use of the LOD graph for enhancing snippets would introduce too much noise. Also, Google Rich Snippet (GRS)~\cite{steiner2010google} is a similar initiative that relies exclusively on structured metadata authored by the Web pages' publishers.

Moreover, a study made in 2012\cite{bizer2013deployment} on the over 40 million websites of the Common Crawl corpus\footnote{\url{http://commoncrawl.org}} shows that 5.64\% of the websites contained embedded structured data. However, nearly 50\% of the top 10,000 websites of the Alexa list of popular websites\footnote{\url{http://www.alexa.com/topsites}} had structured data. Moreover, the authors of the study say that: ``The topics of the data [\ldots] seem to be largely determined by the major consumers the data is targeted at: Google, Facebook, Yahoo!, and Bing''. Therefore, there is still a clear need for a high quality process that, given a document relevant to a Web search query, can select the most relevant resources among those automatically discovered within the document (e.g., through state of the art NLP algorithms), and this, whatever the document's provenance may be. An efficient algorithm for ranking the resources of a LOD graph while taking into account their textual context could serve this purpose.

However, most of the existing approaches that can be used to rank the resources of graphs coming from the Web of data are not well adapted to this task. Thus, OntologyRank~\cite{ding2004swoogle} (used by Swoogle) introduces a modified version of PageRank with a teleportation matrix that takes into account the types of the links between ontologies. Similarly, PopRank~\cite{nie2005object} offers a modified PageRank that considers the different types of predicates between resources. RareRank~\cite{wei2011rational} introduces a modified PageRank with a teleportation matrix that takes into account topical relationships between resources as available from ontologies. The approach introduced in~\cite{fafalios2014post} modifies the teleportation matrix by taking into account the ranking of the Web pages within which the resources were discovered. Since this approach can be applied to our context, we include it to our evaluations (see section~\ref{eval}). Finally, TRank~\cite{tonon2013trank} addresses the task of ranking entity types given an initial entity and its textual context.

Given this context, we introduce LDRANK, a query-biased and context-aware ranking algorithm for LOD resources. Moreover, we apply LDRANK to the construction of generic semantic snippets that can apply to any Web page. In the next section, we introduce how we built a dataset through crowdsourced relevance judgments to evaluate our algorithm, LDRANK.

\section{Dataset for Evaluating Query-biased Ranking of LOD resources}\label{dataset}
We are interested in query-biased algorithms for the ranking of resources in sparse and heterogeneous LOD graphs associated with a textual context. To our knowledge, there is no evaluation dataset suited to this context (this can be verified for example through a recent survey~\cite{roasurvey}). Therefore, we used a crowdsourcing approach for making our evaluation dataset (freely available online\footnote{\url{http://liris.cnrs.fr/drim/projects/ensen/}}). We now describe how this dataset was obtained.

\subsection{Data Collection}
We took randomly 30 queries from the ``Yahoo! Search Query Tiny Sample'' offered by Yahoo! Webscope\footnote{\url{http://webscope.sandbox.yahoo.com/catalog.php?datatype=l}}. We submitted the queries to the Google search engine and we kept the top-5 Web pages for each query. For each one of the 150 HTML Web pages, we extracted its main raw textual content by applying the algorithm proposed by Kohlschütter, Fankhauser, and Nejdl~\cite{kohlschutter2010boilerplate}. On average, the text we kept for each Web page is made of 467 words. We applied DBpedia Spotlight~\cite{mendes2011dbpedia} on these texts to detect resources. There are on average 81 detected resources by Web page.

\subsection{Microtasks Generation}
Considering the length of our texts, the task of evaluating all the annotations of a Web page would be too demanding. Therefore, we divide this task into smaller ``microtasks''. A microtask will consist in scoring the relevance of the annotations of a single sentence. We split the text of a Web page into sentences with the ICU BreakIterator algorithm\footnote{\url{http://icu-project.org/apiref/icu4c/classicu\_1\_1BreakIterator.html}}. There are on average 22 sentences by document. Moreover, if a sentence contains more than 10 annotated resources, the work will be split over multiple microtasks. We used the CrowdFlower\footnote{\url{http://www.crowdflower.com/}} crowdsourcing platform. It distributes work to contributors in the U.S. and 153 other countries while maintaining quality and controlling costs. It has a global pool of 5 million contributors. A microtask is called a job by CrowdFlower. The design of a job is specified in CML, a markup language provided by CrowdFlower. For each job, we give the worker a short list of instructions about how to complete the job (we tested many formulations until finding a suitable one understood by all workers). We provide the worker with a topic made of a title (the query) and a short text (the sentence). For each resource in the sentence, there is a question asking the worker to evaluate the correctness and the relevance of the annotation. We used the ordinal scale proposed by J{\"a}rvelin and Kek{\"a}l{\"a}inen when they introduced the DCG graded relevance\cite{jarvelin2000ir}: irrelevant (0), marginally relevant (1), fairly relevant (2), and highly relevant (3). Each question is associated with a small text that describe the resource (viz. the beginning of its DBpedia abstract). Each job was given to 10 workers. Therefore, for each job we have 10 judgments. Each job was paid \$.01.

\subsection{Quality Control}
We only accepted workers that had completed over a hundred questions across a variety of job types and had an high overall accuracy. Workers had a maximum of 30 minutes to provide an answer. Workers had to spend at least 10 seconds on the job before giving an answer. We measured the agreement between workers with the Krippendorff's alpha coefficient~\cite{krippendorff2012content}. This coefficient uses by default a binary distance to compare answers, but other distances can be used. To take into account the fact that we used an ordinal scale encoding both correctness and relevance, we used the following symmetric distance: $d(0, 1)=0.5$ ; $d(0, 2)=0.75$ ; $d(0, 3)=1$ ; $d(1, 2)=0.25$ ; $d(1 ; 3)=0.5$ ; $d(2 ; 3)=0.25$ ; $d(x,x)=0$. With these parameters, we obtained an alpha of $0.22$. According to Landis and Koch's scale~\cite{landis1977measurement}, this can be considered a fair agreement (the scale was designed for Fleiss' kappa, but the Krippendorff's alpha is in most ways compatible with the kappa). However, by comparison with existing works that applied crowdsourcing to an information retrieval context, we cannot be satisfied with an alpha of $0.22$. For example, Jeong~{\it et al.}~\cite{jeong2013crowd} obtained a Fleiss' kappa of $0.41$ (i.e. moderate agreement) for a crowd-powered socially embedded search engine. However, Alonso, Marshall, and Najork~\cite{alonsocrowdsourcing} obtained a Krippendorff's alpha between $0.03$ and $0.19$ for a more subjective task: deciding if a tweet is or is not interesting. To improve the quality of our dataset, we found the workers that often disagreed with the majority. In fact, by removing the workers that disagree with the majority in more than $41.2\%$ of the cases, we obtained a Krippendorff's alpha of $0.46$. Then, $96.5\%$ of the jobs are done by at least 3 workers, $66\%$ of the jobs are done by at least 5 workers, and we have only $0.7 \%$ of the jobs done by only 1 worker.

\subsection{Aggregation of the Results}
We used majority voting for aggregating the results within each sentence. We used two different methods to break ties : (i) the maximum of the mean of the workers' trust (a metric provided by CrowdFlower), or (ii) the highest value. We discovered later that these two choices result in very similar outcomes when the dataset is used to compare ranking algorithms. We used the same majority voting strategy to aggregate the results at the level of a Web page.

In the next section, we introduce LDRANK, a query-biased ranking algorithm for LOD resources. The dataset we just described will be used in section~\ref{eval} to evaluate LDRANK and to compare it to the state of the art.
\section{LDRANK, a Query-biased Ranking Algorithm for LOD Resources}\label{ldrank}
\subsection{Context}
We introduce LDRANK (Linked Data Ranking Algorithm), a quey-biased algorithm for ranking the resources of a RDF graph. We suppose that the resources were discovered in a Web page found by a Web search engine in answer to a user's query.

In our experiments, the resources are detected in the Web page by DBpedia Spotlight~\cite{mendes2011dbpedia}. From this set of resources and through queries to a DBpedia SPARQL endpoint, we obtain a graph by finding all the relationships between the resources. To each resource, we associate a text obtained by merging its DBpedia abstract and windows of text (300 characters) from the Web page centered on the surface forms associated with the resource. We remove the empty words and we apply stemming\footnote{\url{http://snowball.tartarus.org}} to this text.

LDRANK is adapted by design to such sparse graphs of LOD resources detected in a Web page.
First, LDRANK uses the explicit structure of the graph through a PageRank-like algorithm; second, it uses the implicit relationships that can be inferred from the text associated with the resources through an original variation of the Singular Value Decomposition (SVD); and third, it takes into account the ranking of the Web pages where the resources were found thanks to a scoring function first introduced by Fafalios and Tzitzikas~\cite{fafalios2014post}. 

More precisely, the SVD-based textual analysis and the exploitation of the ranking obtained from a Web search engine result page, each produce a different probability vector expressing some prior knowledge (or belief) about the importance of the resources (see sections~\ref{srp} and~\ref{svd}). Next, these probability vectors are combined through a consensual linear opinion aggregation strategy first introduced by Carvalho and Larson~\cite{carvalho2013consensual} (see section~\ref{belief}). Finally, we use this combined prior knowledge to influence the convergence of a PageRank-like algorithm towards a stable probability distribution corresponding to the final ranking of the resources (see section~\ref{mainalgo}).

\subsection{Prior Knowledge Based on a Web Search Engine Result Page}\label{srp}

\algbegin Algorithm H (Hit Score). This algorithm computes a probability vector ($hitdistrib$) that represents prior knowledge about the importance of the resources based on the rank of the Web pages in which they were detected. This strategy was first introduced by Fafalios and Tzitzikas~\cite{fafalios2014post}.
\algstep H1. $A \leftarrow $ the list of the top Web pages ranked by a Web search engine.
\algstep H2. $E \leftarrow $ the set of detected resources.
\algstep H3. $docs(e) \equiv $ the documents of $A$ containing the detected resources $e$.
\algstep H4. $rank(a) \equiv $ the rank of document $a$ in $A$.
\algstep H5. $hitscore(e) \equiv \sum_{a \in docs(e)} (size(A) + 1) - rank(a)$
\algstep H6. $hitdistrib[e] \leftarrow hitscore(e) / \sum_{e' \in E} hitscore(e')$
\algstep H7. [{\it End.}] \quad\slug

\subsection{Prior Knowledge Based on a Latent Analysis of Textual Data}\label{svd}

\aalgbegin Algorithm S (Linked Data Iterative SVD). This algorithm computes a probability vector ($svddistrib$) that represents prior knowledge about the importance of the resources based on the textual data associated to them.
\algstep S1. [{\it Initial matrix.}] $R \leftarrow $ the sparse resource-stem matrix (i.e., resources in rows, stems in columns) in Compressed Column Storage (CCS) format\footnote{\url{http://netlib.org/linalg/html\_templates/node92.html}}.
\algstep S2. [{\it Initial important resources.}] $info\_need \leftarrow $ a set of resources made of the union of the resources detected in the text of the query and the one resource with the best hitscore (for the case when no resources were detected in the query). We assume that these resources are likely to be close to the information need of the user.
\algstep S3. [{\it First SVD.}] $(U,S,V^T) \leftarrow svdLAS2A(R, nb\_dim)$ Compute the singular value decomposition (SVD) of $R$ at rank $k = nb\_dim$. Since $R$ is very sparse, we use the {\it las2} algorithm developed by Michael W.  Berry~\cite{berry1992large} to compute the decomposition: $R_k = U_k S_k V_k^T$ with $U_k$ and $V_k$ orthogonal, $S_k$ diagonal, such that $\|R-R_k\|_F$ is minimized (i.e. from the perspective of the Frobenius norm, $R_k$ is the best rank-$k$ approximation of $R$).
\algstep S4. [{\it Resources' coordinates in the reduced space.}] $SUT \leftarrow S U^T$ In the new $k$-dimensional space, this operation scales the coordinates of the resources (i.e. the rows of $U$) by their corresponding factor in $S$. This is done by the matrix product: $S U^T$. Thus, we obtain the coordinates of the resources in the reduced space (i.e. the columns of $SUT$).
\algstep S5. $prev\_norms \leftarrow $ euclidean norms of the resources in the reduced space.
\algstep S6. [{\it Updated matrix.}] $R' \leftarrow $ $R$ where the rows corresponding to the resources of $info\_need$ have been multiplied by the parameter $stress$ (since $R$ is in CCS format, it is more convenient to do this operation on the transpose of $R$).
\algstep S7. [{\it Second SVD.}] $(U',S',V'^T) \leftarrow svdLAS2A(R', nb\_dim)$
\algstep S8. [{\it Updated resources' coordinates in the reduced space.}] $SUT' \leftarrow S' U'^T$
\algstep S9. $norms \leftarrow $ updated euclidean norm of the resources in the reduced space.
\algstep S10. [{\it Drift of the resources away from the origin of the reduced space}.] \\$svdscore(e) \equiv norms[e] - prev\_norms[e]$.
\algstep S11. $svddistrib[e] \leftarrow svdscore(e) / \sum_{e'} svdscore(e')$
\algstep S12. [{\it End.}] \quad\slug

We shall now introduce the essential property of the SVD on which relies Algorithm~S. For a strong dimensional reduction (i.e. for small values of $k$), the transformation $S_k U^T$ tends to place resources that were orthogonal to many other resources in the row space of $R$ near the origin of the $k$-dimensional resulting space. Indeed, as we said above, the SVD can be seen as an
optimization algorithm, and to minimize the error due to the impossibility for a resource to be orthogonal to more than $k$ non co-linear resources, this resource should be placed as close to the origin as possible for its dot product with other resources to remain small. A similar argument can be used to show that resources co-linear to many other resources in the row space of $R$ will also tend to be near the origin of the $k$-dimensional space.

Algorithm~S uses this property for ranking the resources by their importance relatively to the user's information need. In $R'$ the resources that are believed to be close to the information need are given artificially more importance. Therefore, resources having interesting relationships with the resources artificially pushed away from the origin will also move away from the origin. By ``interesting'', we mean different from the relationships they maintain with much of the other resources (cf. the geometric argument developed above about the SVD seen as an optimization algorithm).

We obtained the best experimental results with a reduction to the 1 dimensional line (i.e. with $nb\_dim = 1$ in steps S3 and S7 of Algorithm~S), and with a stress factor (step S6 of Algorithm~S) of $1000.0$.

\subsection{Belief Aggregation Strategy}\label{belief}

We consider $hitdistrib$ (from Algorithm~H), $svddistrib$ (from Algorithm~S), and the equiprobable distribution ($equidistrib$) as three experts' beliefs (or prior knowledge) about the importance of the resources. To aggregate these beliefs, we apply Carvalho and Larson~\cite{carvalho2013consensual} consensual linear opinion pool algorithm. It is an iterative algorithm where at each step expert $i$ re-evaluates its distribution as a linear combination of the distributions of all the experts. The weight associated by expert $i$ to the distribution of expert $j$ is proportional to the distance between the two distributions. The authors define this distance such that the process converges towards a consensus. We will refer to this resulting consensual probability vector by the name $finaldistrib$.

\subsection{LDRANK}\label{mainalgo}

The PageRank~\cite{page1999pagerank} algorithm transforms the adjacency matrix ($M$) of a network of Web pages into a matrix $H$ which is both stochastic (i.e., each row of $H$ sums to 1) and primitive (i.e., $\exists k$ s.t. $H^k > 0$), thus assuring the existence of a stationary vector (i.e., the positive eigenvector corresponding to the eigenvalue $1$). This stationary vector is a probability vector that can been interpreted as representing the importance of each Web page. Moreover, it can be computed efficiently with the power iteration algorithm by taking into account the sparsity of the stochastic matrix.

In the original version of the PageRank algorithm, no assumption is made about the probability of importance of the Web pages before the link analysis takes place. In other words: first, the matrix $M$ is transformed into a stochastic matrix $S$ by replacing each null row by the equiprobable distribution ($equidistrib$); second, the matrix $S$ is transformed into a primitive matrix $H$ by a linear combination with the so-called teleportation matrix ($T$): $H = \alpha S + (1-\alpha)T$ where each row of $T$ is the equiprobable distribution ($equidistrib$).

In algorithm~LDRANK, instead of using the equiprobable distribution, we use the consensual distribution ($finaldistrib$) introduced above in section~\ref{belief}. We obtained the best experimental results for $0.6 \leq \alpha \leq 0.8$. Moreover, we set at $1E-10$ the value of the convergence threshold controlling the termination of the power iteration method that computes the stationary vector.

LDRANK is available online under an open-source license\footnote{Source code available online under an opensource license \url{http://liris.cnrs.fr/drim/projects/ensen/}}.

\subsection{LDRANK evaluation}\label{eval}
We compared four ranking strategies, each one of them is based on a different source of prior knowledge used to inform a PageRank-like algorithm: unmodified PageRank i.e., prior knowledge about the importance of the resources is modeled by an equiprobable distribution (we name this strategy EQUI); PageRank modified with the hitscore prior knowledge introduced in section~\ref{srp} and due to  Fafalios and Tzitzikas~\cite{fafalios2014post} (named HIT); PageRank modified with our new SVD-based prior knowledge introduced in section~\ref{svd} (named SVD); and PageRank modified with a consensual mixture of the three previous sources of prior knowledge (named LDRANK).

In order to compare the four strategies (EQUI, HIT, SVD and LDRANK), we used the NDCG (Normalized Discounted Cumulative Gain) metric. The DCG (Discounted Cumulative Gain) at rank $r$ is defined as: $DCG_r = rel_1 + \sum_{i=1}^r \frac{rel_i}{log_2 i}$. NDCG at rank $r$ is DCG at rank $r$ normalized by the ideal ranking at rank $r$. The construction of the dataset used for the evaluation was introduced in section~\ref{dataset}. 

\begin{figure}
\centering
\includegraphics[width=4in, keepaspectratio]{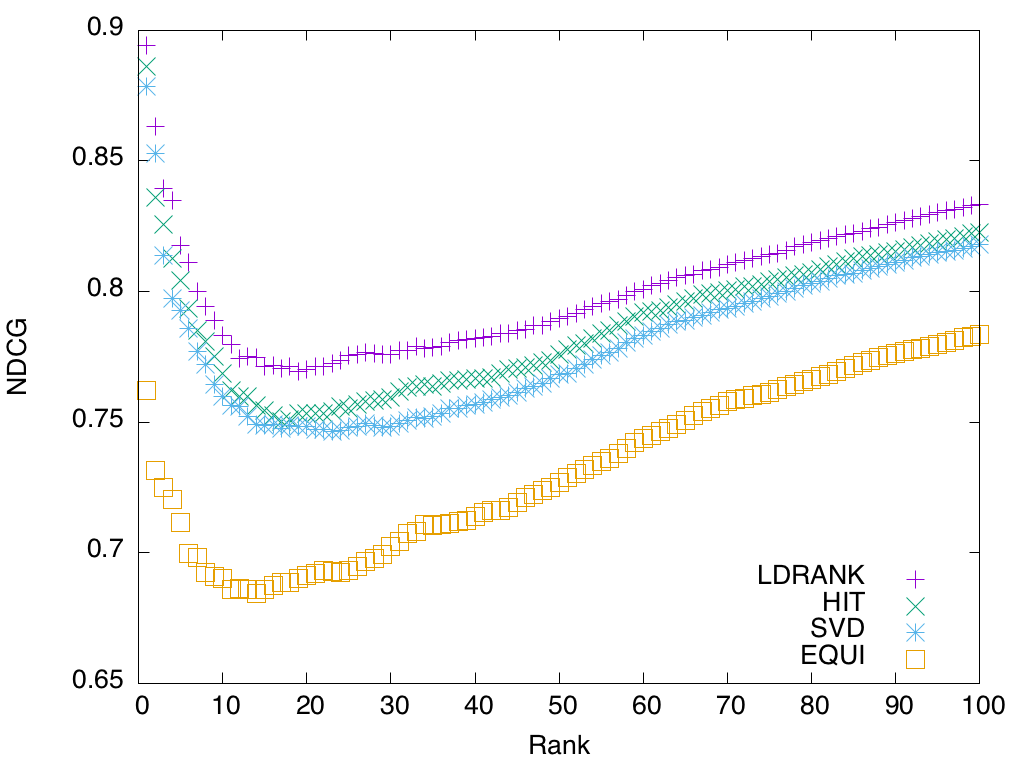}
\caption{Comparison of the NDCG scores for the four different strategies}
\label{fig:ndcg}
\end{figure}

\begin{figure}
\centering
\includegraphics[width=4in, keepaspectratio]{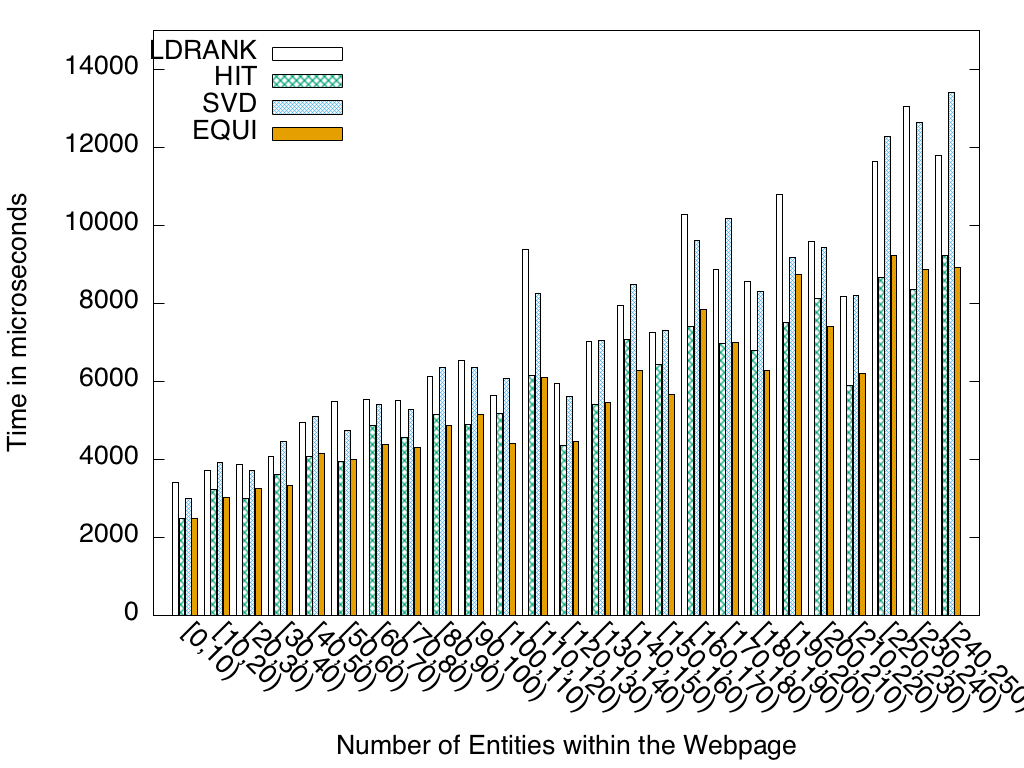}
\caption{Comparison of the execution time for the four different strategies (with processor: 2.9~GHz Intel Core i7, and memory: 8~GB 1600~MHz DDR3)}
\label{fig:perfs}
\end{figure}

The results are presented in Figure~\ref{fig:ndcg}. We can see that the SVD and HIT strategies obtain similar performances. However, they are clearly outperformed by their consensual combination. Moreover, since we systematically took into account the sparsity of the data, we obtain good execution time performances (see Figure~\ref{fig:perfs}). The SVD strategy takes more time than the HIT strategy since it needs to compute the SVD. The additional time spent by the combined strategy is due to the time necessary to converge towards a consensus. Finally, we did similar experiments by considering the edges of the graph bidirectional. The relative performance and accuracy of the algorithms were similar, but the absolute NDCG scores were slightly better.

It should be noted that through these experiments, beside introducing a new efficient ranking strategy based on an original use of the SVD dimensionality reduction, we are also offering evidence that different strategies based on a modification of the teleportation matrix of the PageRank algorithm can profitably be combined when considered as concurrent sources of prior knowledge about the importance of the resources.

\section{Overview of ENsEN}\label{ensen}
In order to better convince the reader of the usefulness and efficiency of LDRANK, we used it at the core of ENsEN (Enhanced Search Engine): a software system that enhances a SERP with semantic snippets (a live demonstration is available online, see a previous footnote for the URL). 
Given the query, we obtain the SERP (we used Google for our experiments). For each result of the SERP, we use DBpedia Spotlight to obtain a set of DBpedia resources. In the same way, we find resources from the terms of the query. From this set of resources and through queries to a DBpedia SPARQL endpoint, we obtain a graph by finding all the relationships between the resources. To each resource, we associate a text obtained by merging its DBpedia's abstract and windows of text from the Web page centered on the surface forms associated with the resource. With as input the graph, its associated text, and the resources extracted from the query, we execute LDRANK and we obtain a ranking of the resources. The top-ranked resources (viz. ``main-resources'') are displayed on the snippet. From a DBpedia SPARQL endpoint, we do a 1-hop extension of the main-resources in order to increase the number of triples among which we will then search for the more important ones.  To do this, we build a 3-way tensor from the extended graph: each predicate corresponds to an horizontal slice that represents the adjacency matrix for the restriction of the graph to this predicate. We compute the PARAFAC decomposition of the tensor into a sum of factors (rank-one three-way tensors) and interpret it in manner similar to~\cite{franz2009triplerank}: for each main-resource, we select the factors to which it contributes the most (as a subject or as an object), and for each one of these factors we select the triples with the best ranked predicates. Thus, we associate to each main-resource a set of triples that will appear within its description. Finally, we used a machine learning approach to select short excerpts of the Web page to be part of the description of each main-resource. In the context of this paper, for lack of space, we cannot describe this process but full details are available in an online technical report (see a previous footnote for the URL).

\section{Crowdsourcing-based User Evaluation}\label{user}
\begin{figure}
\centering
\includegraphics[width=4.9in, keepaspectratio]{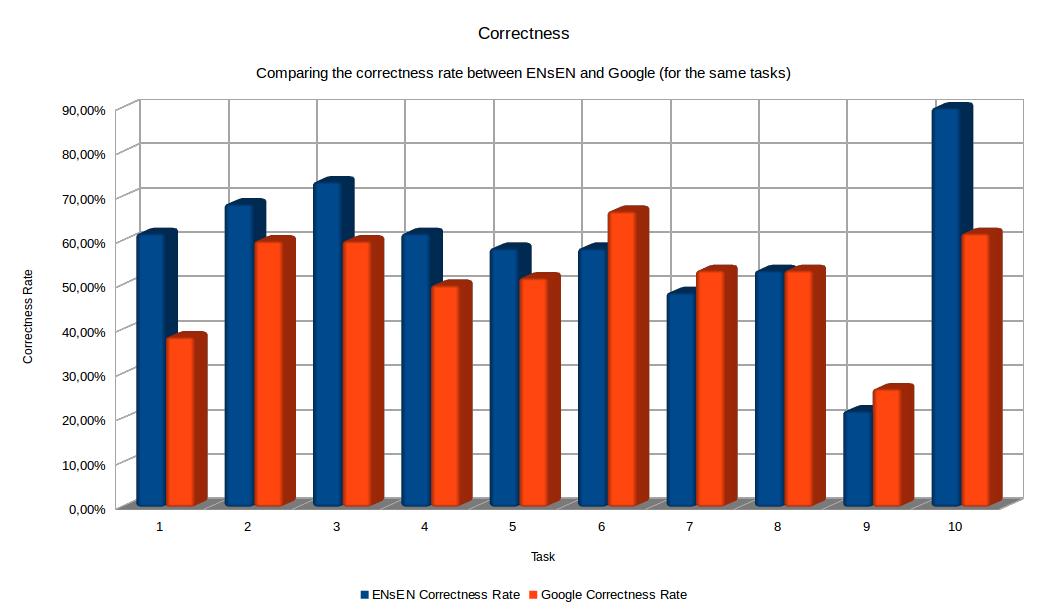}
\caption{Average Number of Correct Answers}
\label{fig:Correctness}
\end{figure}

\begin{figure}
\centering
\includegraphics[width=4.9in, keepaspectratio]{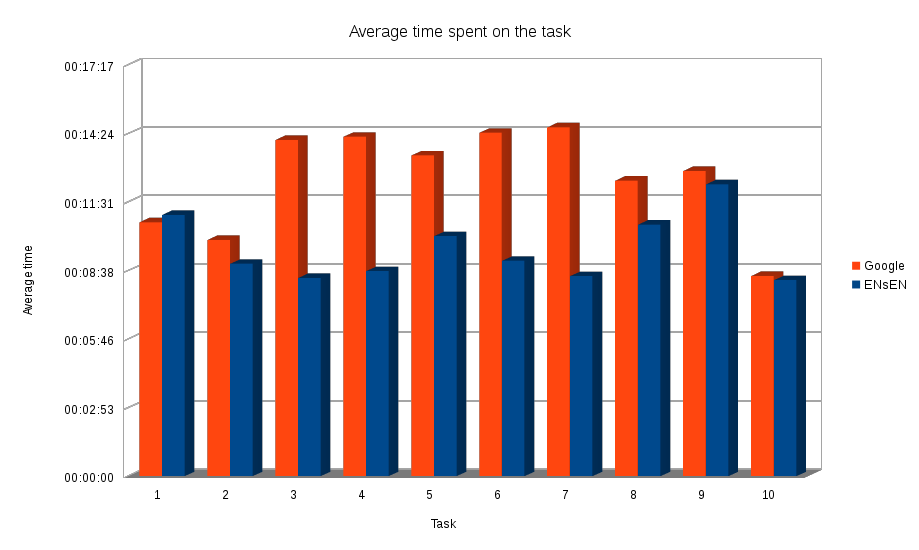}
\caption{Time Spent for Answering the Tasks}
\label{fig:Timespent}
\end{figure}

We selected randomly 10 tasks from the ``Yahoo! Answers Query To Questions'' dataset\footnote{\url{http://webscope.sandbox.yahoo.com/catalog.php?datatype=l}}. Each task was made of three questions on a common topic. To each task corresponds a job on the CrowdFlower platform. Each job was priced \$0.20. We collected 20 judgments for each task. Half of the workers was asked to use our system, and the other half used Google. In order to control that a worker answered the task by using our system, we generated a code that the worker had to copy and paste into her answer. The correctness results are shown on Figure~\ref{fig:Correctness}. Only complete answers were considered correct. We also monitored the time spent to answer the tasks (see Figure~\ref{fig:Timespent}). Thus, ENsEN is clearly beneficial to its users in terms of usefulness.

\section{Conclusion}\label{conclusion}
We proposed a new algorithm, LDRANK, for ranking the resources of a sparse LOD RDF graph given the knowledge of a user's information need expressed as a query made of keywords. These kind of graphs appear in particular as the result of the automatic detection of resources in a Web page. LDRANK takes advantage of both the explicit structure given by the Web of data and the implicit relationships that can be found by text analysis of a Web page. We applied LDRANK in the context of semantic snippets where its high accuracy allowed for the construction of useful and usable enhanced snippets that integrate resources obtained from the automatic annotation of a Web page. Future work could evaluate the potential of this approach for exploratory search.
\bibliographystyle{splncs03}
\bibliography{ESWC2015}
\end{document}